# Role of Square Planar Coordination in the Magnetic Properties of Na$_4$IrO$_4$

Xing Ming,[1,2] Carmine Autieri,[2] Kunihiko Yamauchi,[3] Silvia Picozzi[2]*

1. College of Science, Guilin University of Technology, Guilin 541004, PR China

2. Consiglio Nazionale delle Ricerche CNR-SPIN, UOS L'Aquila, Sede Temporanea di Chieti, 66100 Chieti, Italy

3. ISIR-SANKEN, Osaka University, 8-1 Mihogaoka, Ibaraki, Osaka, 567-0047, Japan

**ABSTRACT**

Iridates supply fertile grounds for unconventional phenomena and exotic electronic phases. With respect to well-studied octahedrally-coordinated iridates, we pay our attention to a rather unexplored iridate, Na$_4$IrO$_4$, showing an unusual square-planar coordination. The latter is key to rationalize the electronic structure and magnetic property of Na$_4$IrO$_4$, which is here explored by first-principles density functional theory and Monte Carlo simulations. Due to the uncommon square-planar crystal field, Ir $5d$ states adopt intermediate-spin state with double occupation of $d_{z^2}$ orbital, leading to a sizable local spin moment, at variance with many other iridates. The square-planar crystal field splitting is also crucial in opening a robust insulating gap in Na$_4$IrO$_4$, irrespective of the specific magnetic ordering or treatment of electronic correlations. Spin-orbit coupling plays a minor role in shaping the electronic structure, but leads to a strong magnetocrystalline anisotropy. The easy axis perpendicular to the IrO$_4$ plaquette, well explained using perturbation theory, is again closely related to the square-planar coordination. Finally, the large single-ion anisotropy suppresses the spin frustration and stabilizes a collinear antiferromagnetic long-range magnetic ordering, as confirmed by Monte Carlo simulations predicting a quite low Néel temperature, expected from almost isolated IrO$_4$ square-planar units as crystalline building blocks.

# I. INTRODUCTION

Recently, 5$d$ Ir oxides (iridates) have attracted extensive attentions, due to the delicate competition between the on-site Coulomb correlation $U$, Hund's coupling $J_H$, spin-orbit coupling (SOC) and crystal field splitting [1, 2, 3, 4]. New phases, emerging phenomena and fascinating physical properties have been uncovered for iridates. For example, Ir-based pyrochlores display a strong enhancement of SOC by correlations, changing from topological band insulator into topological Mott insulator [1], and orthorhombic perovskite iridates AIrO$_3$ (A = alkaline-earth metal) is proposed as a new class of topological crystalline metals [4]. Most of these studies focused on tetravalent (Ir$^{4+}$, 5$d^5$) iridates, sharing IrO$_6$ octahedron as a common crystalline basis block, where the 5$d$ states are split into triply degenerate $t_{2g}$ states and doubly degenerate $e_g$ states by the octahedral crystal field (see Figure 1). As a result of the interplay between SOC and crystal field splitting, the sixfold degenerate (including the spin degree of freedom) Ir $t_{2g}$ states are split into completely filled quartet $J_{eff}$ = 3/2 and half-filled doublet $J_{eff}$ = 1/2 states [5, 6]. The half-filled $J_{eff}$ = 1/2 level with a hole state is proposed to be a key factor in driving exotic phenomena in iridates [3].

To the best of our knowledge, Ir atoms in iridates have been almost exclusively bonded to oxygen atoms in the form of octahedra. On the contrary, Na$_4$IrO$_4$ features one of the few examples of square-planar coordination geometry in iridates (Figure 1), composed by loosely connected IrO$_4$ square-planar plaquettes [7, 8]. The isolated square-planar IrO$_4$ plaquettes locate in the $ab$ plane with tiny deviations of the Ir-O bonds from the crystallographic $a/b$ axis. For each Ir atom, there are two nearest-neighbor (NN) Ir atoms along the $c$ axis and eight next-nearest-neighbor (NNN) Ir atoms along the lattice diagonal. Na$_4$IrO$_4$ can therefore be viewed as consisting of rigid IrO$_4$ clusters almost separated one from the other, arranged on a body-centered tetragonal lattice. The most remarkable feature in Na$_4$IrO$_4$ is therefore the uncommon local geometry of IrO$_4$ plaquette, which will lead the $d$ orbitals to further split under a square-planar crystal field (as schematically shown in Figure 1). The square-planar geometry is frequently found in 3$d$ transition metal compounds, such as the infinite-layer cuprates SrCuO$_2$ and CaCuO$_2$ or iron oxide SrFeO$_2$ [10]. However, square-planar units are corner-sharing in 3$d$ compounds [10], whereas IrO$_4$ square-planar units are separated one from the other in Na$_4$IrO$_4$ [8]. One of the few experimental investigations on Na$_4$IrO$_4$ showed a temperature dependence of the magnetic susceptibility exhibiting clear antiferromagnetic (AFM) ordering at 25 K [8], although the detailed magnetic structure was not reported. First-principles density functional theory (DFT) calculations proposed the crucial role of effective Coulomb interactions (Hubbard $U$) in determining the crystal structure of $Na_4IrO_4$. In contrast, the magnetic ordering and SOC was reported to play almost no role in the crystal field splitting, orbital filling and structural instability of $Na_4IrO_4$ [8].

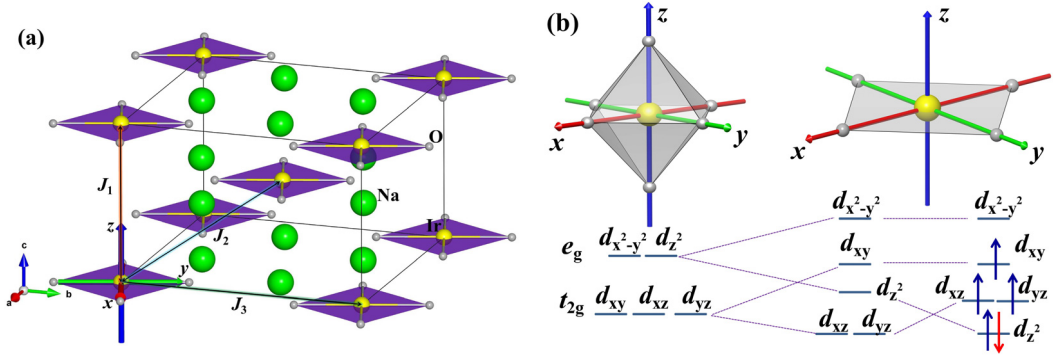

**Figure 1** (a) Crystal structure and spin exchange paths ($J_1$, $J_2$ and $J_3$) of $Na_4IrO_4$. The large (green), middle (yellow), and small (grey) spheres represent the Na, Ir, and O ions, respectively. We use *xyz* for the local coordinates and *abc* for the global orientation. (b) Schematic *d*-orbital splittings under octahedral (left) and square-planar (middle) crystal field, and the actual (right) orders of the energy level arrangements and the intermediate-spin state for $Ir^{4+}$ ($5d^5$) ions in $Na_4IrO_4$ (see below).

In the present work, we explored the electronic structure, magnetocrystalline anisotropy (MCA) and spin exchange interactions in $Na_4IrO_4$ by performing DFT calculations, complemented by Monte Carlo (MC) simulations to predict the Néel temperature and magnetic ground state. All the remarkable properties of $Na_4IrO_4$ are closely related to the crucial square-planar coordination in $IrO_4$ plaquettes. The electronic structure shows an energy level splitting consistent with square-planar crystal field and with strong hybridizations (both inter-atomic between Ir $5d$ and O $2p$ states as well as intra-atomic between Ir $5d_{z^2}$ and Ir $6s$ states), leading to an intermediate-spin state, quite unusual for iridates but expected from almost isolated square-planar $IrO_4$ units. The insulating band gap originates from the strong crystal field splitting, independently on the magnetic ordering and Coulomb interactions. SOC interactions almost have no effect on the electronic structure, but result in a large easy-axis MCA (single-ion anisotropy (SIA)) of $Ir^{4+}$ ion in the unusual square-planar crystal field. Finally, our MC simulations predict a rather low Néel temperature and a collinear long-range AFM magnetic ordering in Na$_4$IrO$_4$, again expected from loosely connected square-planar IrO$_4$ plaquettes.

## II. COMPUTATIONAL METHODS AND STRUCTURAL DETAILS

Electronic structure calculations were carried out using the Vienna Ab Initio Simulation Package (VASP) code [11] within the projector augmented wave (PAW) method [12, 13]. The generalized gradient approximation (GGA) exchange-correlation functional as parameterized by the Perdew-Burke-Ernzerhof (PBE) was used for all spin polarized calculations [14]. SOC was included in the simulations using the noncollinear magnetism settings. The rotationally invariant + $U$ method introduced by Liechtenstein *et al*. was employed to account for correlations effects [15]. The values of the Coulomb interactions $U$ and the Hund's coupling $J_H$ for Ir 5$d$ orbitals were fixed to 2 and 0.2 eV, respectively. $K$-point meshes of 8 × 8 × 12 for the primitive unit cell and 6 × 6 × 6 for $\sqrt{2} \times \sqrt{2} \times 2$ supercell (see below) were used for the Brillouin zone integration. The cutoff energy was set to 520 eV for all DFT calculations. The threshold for self-consistent-field energy convergence was chosen as 10$^{-6}$ eV.

X-ray crystal structure refinements of Na$_4$IrO$_4$ show a tetragonal structure (space group *I4/m*) with two formula units (f. u.) per unit cell. According to the symmetry, Na, Ir, and O atoms can be classified as three nonequivalent crystallographic sites in the unit cell. They are located at 8$c$ (x, y, 0), 2$a$ (0, 0, 0), and 8$h$ (x, y, 0) sites, respectively. From x-ray diffraction experiments, the lattice constants of Na$_4$IrO$_4$ were determined to be $a = b = 7.184$ Å and $c = 4.725$ Å [8]. In the IrO$_4$ square-plane, there are two O-Ir-O bonds, which are mutually perpendicular but slightly deviating from the global crystallographic *a*/*b* axis in the *ab* plane. To monitor the behavior of the square-planar crystal field, a local coordinate system (x , y , z ) defined in Figure 1 is employed for Ir atoms, with z being exactly perpendicular to the IrO$_4$ square-plane, and x, y are defined exactly along one of the Ir-O bonds in the square-planar IrO$_4$ plaquette.

Based on experimental lattice parameters, we optimized all independent atomic internal coordinates and lattice constants. As the detailed magnetic structure is not available [8], the AFM ordering has been simulated by considering an antiparallel alignment of the spin magnetic moment of two Ir atoms in the unit cell, found from first-principles to be the lowest-energy magnetic state (see below). We confirmed that a reasonable $U$ parameter and SOC have only a small impact on the crystal structure. As listed in Table I, our theoretical calculated lattice parameters were in good agreement with available experimental and theoretical results, with errors less than 2% for the lattice constants and 4% for the volume. We noted that smaller errors and similar results to ref. 8 can be obtained for a nonmagnetic state setting. Electronic structure calculations were carried out with the relaxed lattice parameters for the AFM state. First, we performed spin polarized calculations within GGA, then took Coulomb interactions $U$ and SOC into account by GGA + $U$, GGA + SOC and GGA + SOC + $U$ calculations. For the SOC calculations, the quantization axis was set along [0 0 1] (the crystallographic $c$ axis, except where specifically noted otherwise).

**TABLE I** Theoretical calculated and experimental measured lattice constants (Å), unit cell volume (V, Å$^3$), atomic internal coordinates and Ir-O bond length (Å) of Na$_4$IrO$_4$.

|  | $a = b$ | $c$ | V | Na | | O | | Ir-O |
|---|---|---|---|---|---|---|---|---|
|  |  |  |  | x | y | x | y |  |
| Exp. [a] | 7.167 | 4.713 | 242.09 | 0.1962 | 0.4059 | 0.2526 | 0.0815 | 1.902 |
| Exp. [b] | 7.184 | 4.725 | 243.85 | - | - | - | - | 1.942 |
| Theo.[b] | 7.207 | 4.704 | 244.33 | - | - | - | - | 1.938 |
| GGA [c] | 7.256 | 4.759 | 250.53 | 0.1974 | 0.4049 | 0.2538 | 0.0826 | 1.937 |
| GGA+$U$ [c] | 7.280 | 4.742 | 251.33 | 0.1966 | 0.4028 | 0.2525 | 0.0819 | 1.932 |
| GGA+SOC [c] | 7.262 | 4.755 | 250.78 | 0.1972 | 0.4043 | 0.2536 | 0.0824 | 1.937 |
| GGA+SOC+$U$ [c] | 7.287 | 4.739 | 251.63 | 0.1965 | 0.4024 | 0.2524 | 0.0818 | 1.933 |

a. ref. 7

b. ref. 8

c. present work

### III. RESULTS AND DISCUSSIONS

#### A. Electronic structure and local magnetic moments

As shown in Figure 2, the band structures show strong localized and flat-band character around the Fermi level ($E_F$), indicating weak interactions because of the loosely connected crystal structure (Figure 1(a)). Unusual for iridates, an insulating gap has opened up even without Coulomb interaction corrections for the AFM state (Figure 2 (a)). Upon inclusion of Coulomb interactions, the band gap remarkably increases, although the essential characteristics of the band dispersion are unaffected (Figure 2(b)). Furthermore, as presented in Figures 2 (c) and (d), except for lifting the degeneracy of the $d_{xz, yz}$ bands [8], the main features of the band structures remain unchanged with or without SOC. An insulating energy gap was obtained even assuming ferromagnetic (FM) ordering [8], so that the insulating nature in $Na_4IrO_4$ does not depend on the magnetic ordering state, Coulomb parameters and SOC, rather being essentially determined by the crystal field splitting of the unusual $IrO_4$ square-planar units.

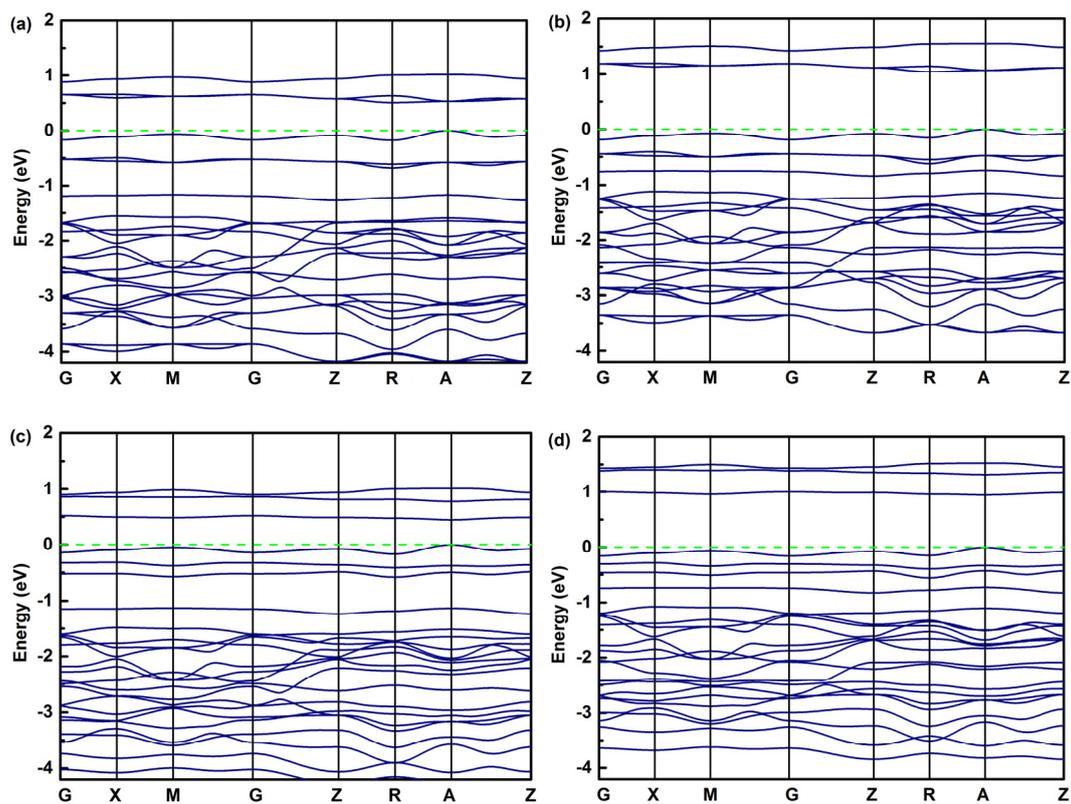

**Figure 2** Band structure of $Na_4IrO_4$ calculated within (a) GGA, (b) GGA + $U$, (c) GGA + SOC, and (d) GGA + SOC + $U$, where $U$ = 2 eV. Since spin up and spin down states are degenerate in the AFM state, only spin up subbands are reported in (a) and (b).

The detailed electronic structure can be further inspected by the projected density of states (pDOS). As shown in Figure 3, due to the inter-atomic interactions between the central Ir and four ligand oxygen ions, the $Ir^{4+}$ 5$d$ states form bonding (ranging from -6 to -4 eV) and antibonding (from -3 to 1 eV) molecular orbitals, with the Ir antibonding states locate around $E_F$ and distinctly split off. This situation is well consistent with the typical energy level splitting of $d$ orbitals under a square-planar crystal field [16], which is so strong that $Ir^{4+}$ ($5d^5$) adopts an intermediate-spin state (see Figure 1 (b)). O $2p$ bands are mainly located in a lower energy range, although the pDOS shows a strong inter-atomic hybridization between Ir $5d$ and O $2p$ states. An insulating gap opens up between different spin channels of spin-up (spin-down) $d_{xy}$ and spin-down (spin-up) $d_{xz,yz}$ orbitals due to the large exchange splitting. The Ir $d_{z^2}$ orbitals are the lowest-lying occupied states for both spin channels. The double occupation of the $d_{z^2}$ orbital (rather than the degenerate $d_{xz, yz}$ orbitals, cfr Figure 1(b) and Figure 4(d)), seemingly at variance with the expectation from crystal field theory for the $D_{4h}$ point symmetry (see the conventional energy level sequence schematically shown in Figure 1(b), bottom middle) [10], also occurred in another infinite-layer $3d$ oxide, $SrFeO_2$, with perfect square-planar coordination [10, 15, 17]. The origin of the $d_{z^2}$-double occupation arises from the reduction of Coulomb repulsion interactions, due to the missing oxygen ions in the direction perpendicular to the $IrO_4$ square-plane (see Figure 1) [18]. In addition, according to the $D_{4h}$ point group symmetry, Ir $5 d_{z^2}$ and $6s$ orbital have the same $a_{1g}$ symmetry, therefore resulting in their intra-atomic hybridization (see pDOS in Figure 3 (c)) and, in turn, to a large reduction of the exchange splitting for the $d_{z^2}$ orbitals and, finally, to their double-occupation [19].

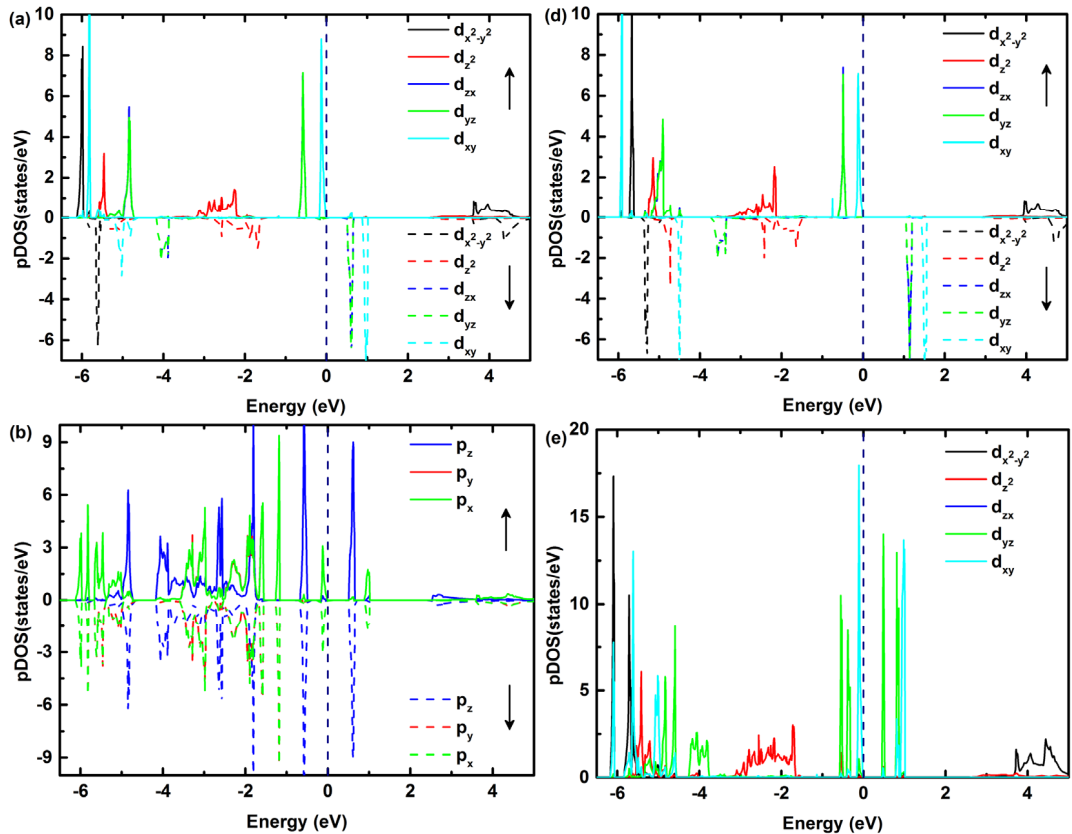

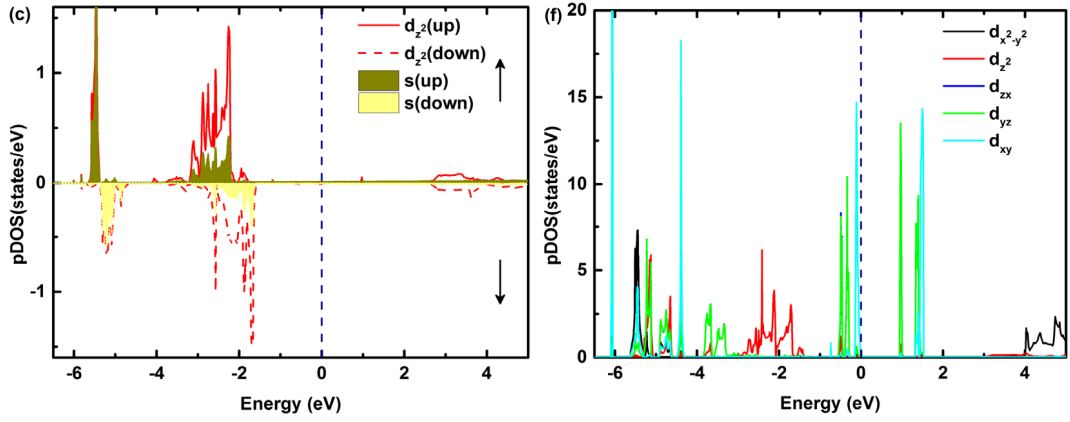

**Figure 3** Projected density of states (pDOS): (a) Ir $5d$, (b) O $2p$ and (c) Ir intra-atomic $5d_{z^2}$ and $6s$ states calculated within GGA; (d), (e) and (f) show Ir $5d$ states calculated within GGA + $U$, GGA + SOC and GGA + SOC + $U$. Due to the structural symmetry, the Ir $5d_{yz}$ and $d_{zx}$ states overlap.

SOC often significantly influences the $5d$ band dispersion and plays an essential role in the insulating ground state for many iridates, with the formation of novel half-filled $J_{eff}$=1/2 spin-orbit insulating states [5, 6, 20, 21]. The $Ir^{4+}$ ions in these iridates all show low-spin $5d^5$ ($t_{2g}^5$, $e_g^0$) electronic configurations. However, as demonstrated in Figure 2, the difference between the band structures with and without SOC is small in $Na_4IrO_4$. In addition, different coordination environments (square-planar vs octahedral) and related crystal fields result in distinct energy level splitting and orbital occupation patterns. As presented in Figure 3, just considering the antibonding states, the $d$-electron configuration in $Na_4IrO_4$ organizes as $(z^2)^1 < (xz, yz)^2 < (xy)^1 < (x^2-y^2)^0$ for the spin-up states, and in sequence $(z^2)^1 < (xz, yz)^0 < (xy)^0 < (x^2-y^2)^0$ for the spin-down states (schematically shown in Figure 1(b)). Due to the strong crystal field splitting, the lowest $d_{z^2}$ states and the highest $d_{x^2-y^2}$ states are located far from other three $d_{xy}$, $d_{yz}$, and $d_{zx}$ states (generally defined as $t_{2g}$ orbitals in octahedral or tetragonal crystal field). In this sense, the electronic configurations of intermediate-spin state $Ir^{4+}$ ions in $Na_4IrO_4$ can be viewed as reduced to a $d^3$ ($t_{2g}^3$, $S = 3/2$) system and the orbital degree of freedom can be thought as being quenched ($L_{eff} = 0$) for a half filled $t_{2g}$ band. According to SOC Hamiltonian $\hat{H}_{SO} = \lambda \hat{S} \cdot \hat{L}$, this could justify why SOC does not play a dominant role in the electronic structure [22, 23, 24, 25].

**TABLE II** Calculated spin moment ($M_S$) and orbital moment ($M_L$) of $Na_4IrO_4$ (values in Bohr

magnetons, positive/negative signs indicate the moment directions).

|  | $M_S$ | | $M_L$ | |
| --- | --- | --- | --- | --- |
|  | Ir | O | Ir | O |
| GGA | ±1.585 | ±0.248 | - | - |
| GGA + $U$ | ±1.744 | ±0.24 | - | - |
| GGA + SOC | ±1.424 | ±0.229 | ∓0.071 | ±0.017 |
| GGA + SOC + $U$ | ±1.592 | ±0.224 | ∓0.045 | ±0.019 |

At variance with other $5d^5$ iridates [5, 26], as shown in Table II, the orbital moment of $Ir^{4+}$ ions is much smaller than its spin moment in $Na_4IrO_4$, indicating that the orbital degree of freedom is indeed quenched and SOC effect is small. In addition, the orbital moments are antiparallel to the spin moments for $Ir^{4+}$ ions, whereas the orbital moments are parallel to the spin moments for $O^{2-}$ ions. These results follow Hund's third rule, according to which the orbital moment and spin moment should be antiparallel (parallel) for a less (more) than half-filled system. Although the calculated spin moment of $Ir^{4+}$ ions is smaller than 3 $\mu B$ for a nominal $S = 3/2$ intermediate-spin $5d^5$-electron system, the spin moment contributions from O atoms are notably large, revealing the strong inter-atomic hybridizations of Ir $5d$ and O $2p$ states, consistent with the pDOS (see Figure 3). A reduced value of spin moment is very common in iridates because of the strong inter-atomic hybridizations between Ir $5d$ and O $2p$ states [5, 25, 27, 28, 29, 30, 31, 32]. However, the spin moments are often smaller than 0.5 $\mu_B$ for $Ir^{4+}$ ions in other octahedral-coordinated iridates [5, 25, 27, 28, 29, 30, 31], whereas the hybridization-driven reduction is far smaller in $Na_4IrO_4$, resulting in large local magnetic moments. At the same time, the orbital moments are often as large as twice of the spin moment in other iridates, where the strong SOC and the large octahedral crystal-field splitting produce an effective $J_{eff}=1/2$ state for the $Ir^{4+}$ ion [5]. The Coulomb interactions are often one order of magnitude smaller in iridates with respect to $3d$-based oxides, and the $5d$ transition-metal oxides are expected to be more itinerant because of the larger spatial extent of $5d$ orbitals [31]. However, the effective electronic correlations increase upon decreasing connectivity of $IrO_6$ octahedra in iridates [33]. Therefore, due to the peculiar crystal structure and square-planar crystal-field splitting, at variance with the expectation from the itinerant of $5d$ iridates, $Na_4IrO_4$ is the only iridate showing an intermediate-spin state with large local spin moments, as demonstrated by the localized flat-band structure and isolated energy levels of $Na_4IrO_4$ (Figure 2).

## B. MCA and preferred spin orientations

**Table III** Calculated MCA energy per M atom (meV) for $Na_4MO_4$ (M = Ru, Rh, Os, and Ir). Total energy values for the spin quantization axis (SAXIS) in the *ab* plane (local [100] and [110] direction) are given with respect to the energy for the SAXIS out of plane (local [001] direction), taken as reference. The SIA energy for $Na_4IrO_4$ with one Ir atom and three non-magnetic Si-ions are given in parentheses.

|                | SAXIS | Ru    | Rh   | Os     | Ir            |
|----------------|-------|-------|------|--------|---------------|
|                | [001] | 0     | 0    | 0      | 0             |
| GGA + SOC      | [100] | -4.87 | 1.80 | -21.09 | 14.88(14.03)  |
|                | [110] | -4.86 | 1.78 | -21.43 | 12.97(12.09)  |
|                | [001] | 0     | 0    | 0      | 0             |
| GGA + SOC + *U*| [100] | -4.34 | 1.65 | -13.17 | 15.99(15.74)  |
|                | [110] | -4.36 | 1.63 | -13.86 | 14.69 (14.44) |

In this paragraph we focus on the MCA in $Na_4IrO_4$ and, for clearer insights, compare it with MCA in other hypothetical 4*d* and 5*d* compounds with square-planar crystal field. Using the initial crystal structure of $Na_4IrO_4$, we replace the Ir ions by $Ru^{4+}$ ($4d^4$), $Rh^{4+}$ ($4d^5$) and $Os^{4+}$ ($5d^4$) ions, respectively. For these hypothetical $Na_4MO_4$ compounds (M = Ru, Rh, and Os), when all the independent atomic internal coordinates and Bravais lattice are allowed to fully relax (including possible relaxation to different space group, coordination, etc), the lattice symmetry of $Na_4IrO_4$ and the related square-planar coordination are kept as ground state (in contrast to what happens for 3*d*-based $Na_4CoO_4$, where the oxygen cage around the 3*d*-metal turns to tetrahedral) [8]. Our calculations predict these $Na_4MO_4$ compounds to show optimized lattice parameters very similar to $Na_4IrO_4$. As reported in Table III, the total energy within GGA + SOC (with or without *U*) strongly depends upon the relative orientation of the spin quantization axis, leading to a sizeable MCA. While this is consistent with the strongly anisotropic coordination in $IrO_4$ "isolated" plaquettes, it would be interesting to experimentally investigate this aspect.

In particular (see Table III), for Na$_4$IrO$_4$ and Na$_4$RhO$_4$ with $d^5$ electronic configurations, the configuration with the spin moments parallel to the $c$ axis (out of plane) is more stable than that with the spin moments in the $ab$ plane. In contrast, for Na$_4$OsO$_4$ and Na$_4$RuO$_4$ with $d^4$ electronic configurations, the states with the spin moments in the $ab$ plane are energetically favored. In other words, the $d^5$ compounds show an easy-axis anisotropy, whereas the $d^4$ compounds show an easy-plane anisotropy.

The MCA and preferred spin orientations can be analyzed via perturbation theory [34, 35, 36, 37], where SOC is included to couple spin and orbital angular momentum ($\hat{S}$ and $\hat{L}$), resulting in the SOC Hamiltonian, $\hat{H}_{SO} = \lambda \hat{S} \cdot \hat{L}$, $\lambda$ being the SOC constant. Employing two independent coordinate systems (x, y, z) and (x', y', z') for the orbital $\hat{L}$ and spin $\hat{S}$, respectively, the SOC Hamiltonian $\hat{H}_{SO} = \lambda \hat{S} \cdot \hat{L}$ is rewritten as $\hat{H}_{SO} = \hat{H}_{SO}^0 + \hat{H}_{SO}'$, where the "spin-conserving" term

$$\begin{aligned}\hat{H}_{SO}^0 &= \lambda \hat{S}_{z'}(\hat{L}_z \cos\theta + \frac{1}{2}\hat{L}_+ e^{-i\phi}\sin\theta + \frac{1}{2}\hat{L}_- e^{+i\phi}\sin\theta)\\ &= \lambda \hat{S}_{z'}(\hat{L}_z \cos\theta + \hat{L}_x \sin\theta\cos\phi + \hat{L}_y \sin\theta\sin\phi)\end{aligned} \quad (1)$$

and the "spin-non-conserving" term

$$\begin{aligned}\hat{H}_{SO}' &= \frac{1}{2}\lambda \hat{S}_{+'}(-\hat{L}_z \cos\theta - \hat{L}_+ e^{-i\phi}\sin^2\frac{\theta}{2} + \hat{L}_- e^{+i\phi}\cos^2\frac{\theta}{2})\\ &+ \frac{1}{2}\lambda \hat{S}_{-'}(-\hat{L}_z \sin\theta + \hat{L}_+ e^{-i\phi}\cos^2\frac{\theta}{2} - \hat{L}_- e^{+i\phi}\sin^2\frac{\theta}{2})\\ &= \frac{1}{2}\lambda(\hat{S}_{+'} + \hat{S}_{-'})(-\hat{L}_z \sin\theta + \hat{L}_x \cos\theta\cos\phi + \hat{L}_y \cos\theta\sin\phi)\end{aligned} \quad (2)$$

where $\theta$ and $\phi$ define the magnetization direction (z') with respect to the (x, y, z) coordinate system [38]. The energy correction by SOC is given by

$$\Delta E_{soc} = \sum_{e,g} \frac{|\langle g|\hat{H}_{soc}|e\rangle|^2}{E_g - E_e} \quad (3)$$

where $|g\rangle$ and $|e\rangle$ are the ground (occupied) and excited (unoccupied) states, $E_g$ and $E_e$ are the corresponding unperturbed energies [33-36, 39, 40].

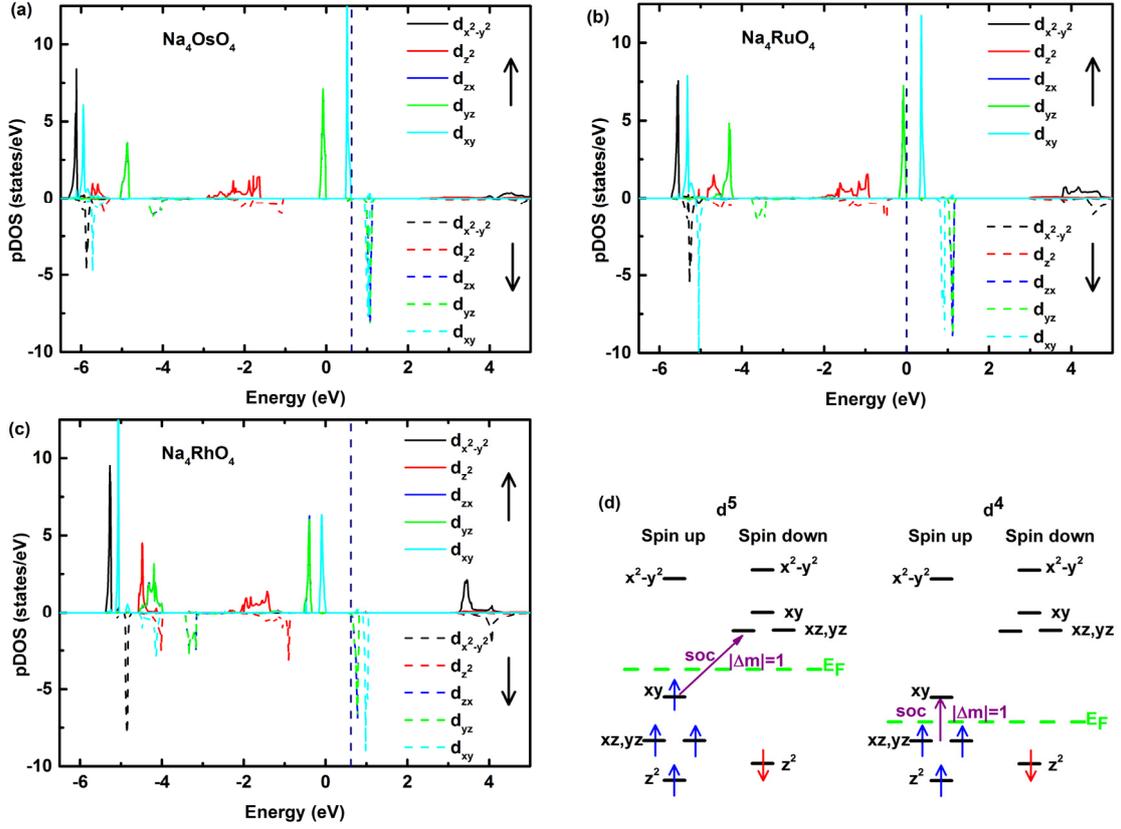

**Figure 4** The pDOS for M $d$ states for Na$_4$MO$_4$ compounds (M = Os$^{4+}$, Ru$^{4+}$ and Rh$^{4+}$): (a) Os $5d$, (b) Ru $4d$ and (c) Rh $4d$ states calculated within GGA. Panel (d) shows the schematic energy level splitting by the square-planar crystal field and the orbital occupations for the $d^5$ and $d^4$ configuration, where the SOC between unperturbed occupied and unoccupied $d$ states is explicitly highlighted.

As shown in the pDOS of Figure 3 and Figure 4, electronic structure calculations indicate that the crystal field splitting is the same for these isostructural 4$d$ and 5$d$ compounds and typical of square-planar splitting [10, 15, 16, 17, 18]. The crystal field splitting is so strong that Ru$^{4+}$ (4$d^4$), Rh$^{4+}$ (4$d^5$), Os$^{4+}$ (5$d^4$) and Ir$^{4+}$ (5$d^5$) ions all adopt intermediate-spin states with double occupation of the $d_{z^2}$ orbitals. Due to the large spin exchange splitting and crystal field splitting, even without Hubbard $U$ corrections, the insulating gaps open up between different (same) spin channels for Na$_4$IrO$_4$ and Na$_4$RhO$_4$ (Na$_4$OsO$_4$ and Na$_4$RuO$_4$) with $d^5$ ($d^4$) electronic configurations.

As schematically shown in Figure 4 (d), with the same $d^5$ electronic configurations, Na$_4$IrO$_4$ and Na$_4$RhO$_4$ display the same energy level splitting and orbital occupations. The smallest energy gap between the occupied and unoccupied levels occurs between the $d_{xy}$ (spin-up) and the $d_{xz, yz}$ (spin-down) levels. These levels differ in their magnetic orbital quantum number $|\Delta m|$ by 1 [34, 35]. Because the occupied and unoccupied $d$ states couple within opposite-spin channels, the SOC Hamiltonian will be governed by spin-non-conserving term $\hat{H}'_{SO}$ (equation (2)), and the perturbation matrix element $\langle g|\hat{H}_{soc}|e\rangle$ will be proportional to $\cos\theta$ [41]. As such, the SOC-induced interactions are maximized when the spin magnetization direction is parallel to the orbital $z$-axis (*i.e.*, $\theta = 0°$) in $Na_4IrO_4$ and $Na_4RhO_4$.

On the other hand, the situation is different when considering $Na_4OsO_4$ and $Na_4RuO_4$ with $d^4$ electronic configurations. As presented in Figure 4, the $d$-electron configurations in $Na_4OsO_4$ and $Na_4RuO_4$ align as $(z^2)^1 < (xz, yz)^2 < (xy)^0 < (x^2-y^2)^0$ for the spin-up states, and in an order of $(z^2)^1 < (xz, yz)^0 < (xy)^0 < (x^2-y^2)^0$ for the spin-down states. According to the schematic energy diagram in Figure 4 (d), the smallest energy gap between the occupied and unoccupied levels occurs now in the same spin-up (or spin-down) channel for $d_{xz,yz}$ and $d_{xy}$ orbitals, differing in their magnetic orbital quantum number $|\Delta m|$ by 1 [34, 35]. SOC interactions couple occupied and unoccupied $d$ states within the same spin channel, so the SOC Hamiltonian will be governed by spin-conserving term $\hat{H}_{SO}$ (equation (1)), and the perturbation matrix element $\langle g|\hat{H}_{soc}|e\rangle$ will be proportional to $\sin\theta$ [40]. In this case, the SOC-induced interactions are maximized when the spin magnetization direction is perpendicular to the orbital $z$-axis (*i.e.*, $\theta = 90°$). Therefore, $Na_4OsO_4$ and $Na_4RuO_4$ show easy-plane anisotropy, in contrast with the easy-axis anisotropy in $Na_4IrO_4$ and $Na_4RhO_4$.

According to previous works [38, 39], we can further evaluate the perturbation matrix element $\langle g|\hat{H}_{soc}|e\rangle$ and hence the energy correction by SOC. Noting that $\langle xy\uparrow|\hat{H}_{soc}|xy\downarrow\rangle = 0$, for the case of $Na_4IrO_4$ and $Na_4RhO_4$ with $d^5$ electronic configurations, the second-order energy shift is given by

$$\Delta E_{soc} = -\lambda^2 \left( \frac{1}{2\Delta_1} + \frac{1}{2\Delta_4} \right) + \lambda^2 \left( \frac{1}{4\Delta_1} - \frac{1}{2\Delta_3} + \frac{1}{4\Delta_4} \right) \sin^2\theta \qquad (4)$$

where $\Delta_1$, $\Delta_3$ and $\Delta_4$ are the energy gaps for the occupied and unoccupied levels between the $d_{xy}$ (spin-up) with $d_{xz, yz}$ (spin-down), $d_{xz, yz}$ (spin-up) with $d_{xz, yz}$ (spin-down) and $d_{xz, yz}$ (spin-up) with $d_{xy}$ (spin-down) orbitals, respectively.

For the case of Na$_4$OsO$_4$ and Na$_4$RuO$_4$ with $d^4$ electronic configurations, the second-order energy shift is given by

$$\Delta E_{soc} = -\frac{\lambda^2}{2\Delta_1} + \lambda^2 \left( -\frac{1}{4\Delta_1} - \frac{1}{2\Delta_3} + \frac{1}{4\Delta_4} \right) \sin^2 \theta \qquad (5)$$

where $\Delta_1$, $\Delta_3$ and $\Delta_4$ are the energy gaps for the occupied and unoccupied levels between the $d_{xz, yz}$ (spin-up) with $d_{xy}$ (spin-up), $d_{xz, yz}$ (spin-up) with $d_{xz, yz}$ (spin-down) and $d_{xz, yz}$ (spin-up) with $d_{xy}$ (spin-down) orbitals, respectively.

As shown in equations (4) and (5), the azimuthal $\phi$ dependence vanished in the perturbation theory up to second order for the energy shift. According to the energy level arrangements and the related energy gaps from the pDOS of Figure 3 and Figure 4, the angle dependent parts of equations (4) and (5) show a $\sin^2 \theta$ dependence with positive/negative values for the Na$_4$MO$_4$ compounds with $d^5$ ($d^4$) electronic configurations, indicating that the magnetization easy axis is out of (resides in) the *ab* plane. We recall that the simple dependence of the total energy on the magnetization angle $\theta$ deduced from second order perturbation theory [38, 42], can be expressed as:

$$E(\theta) - E_0 = K_1 \sin^2 \theta \qquad (6)$$

To carefully evaluate the dependence of total energy on $\theta$, we performed a series of calculations by rotating the magnetization angle $\theta$. As shown in Figure 5, the calculated results fit well with what expected from equation (6) for all the Na$_4$MO$_4$ (M = Ir$^{4+}$, Os$^{4+}$, Ru$^{4+}$ and Rh$^{4+}$) compounds. The MCA energy (MAE) curves display opposite trend for the $d^5$ (Na$_4$IrO$_4$ and Na$_4$RhO$_4$) and $d^4$ (Na$_4$OsO$_4$ and Na$_4$RuO$_4$) compounds, consistently with the opposite sign of equations (4) and (5) for the angle dependent parts of the energy corrections by SOC. The MAE behavior (Figure 5 (a)) shows a minimum for Na$_4$IrO$_4$ and Na$_4$RhO$_4$ at magnetization direction along the crystallographic *c* axis, corresponding to the easy-axis anisotropy (*i.e.*, the $\theta = 0°$ spin orientation) of $d^5$ materials. In contrast, the MAE behavior (Figure 5 (b)) displays a minimum for Na$_4$OsO$_4$ and Na$_4$RuO$_4$ for magnetization perpendicular to the *c* axis, in line with the easy-plane anisotropy (*i.e.*, the $\theta = 90°$ spin orientation) of $d^4$ materials.

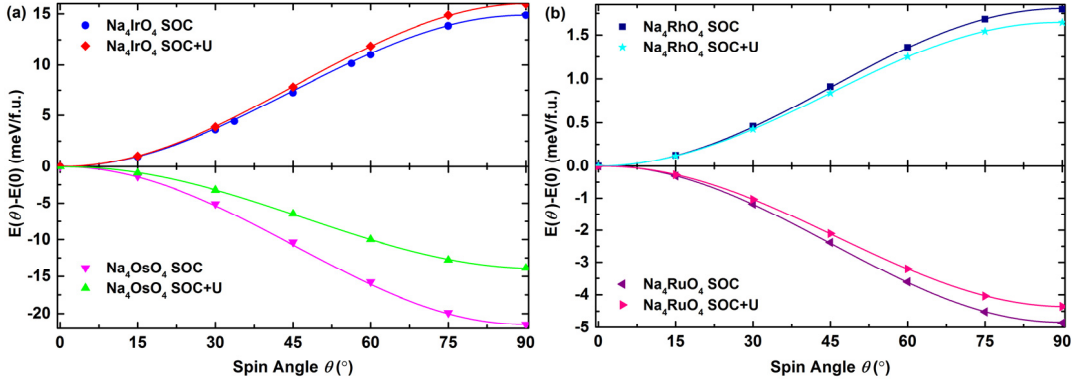

**Figure 5** Dependence of the total energy on the magnetization angle $\theta$ for the Na$_4$MO$_4$ (M = Ir$^{4+}$, Os$^{4+}$, Ru$^{4+}$ and Rh$^{4+}$) compounds and fitted with a function, like $A\sin^2\theta$ (the line). The calculations were performed within GGA including SOC (with or without $U$).

### C. Single-ion anisotropy and spin exchange interactions

From a general point of view, SOC can lead to inter-site Dzyaloshinskii-Moriya (DM) (antisymmetric) interaction, to anisotropic exchange and to single-ion (or single-site) anisotropy (SIA). According to the crystal symmetry, the DM interaction should not appear, due to the presence of inversion symmetry in Na$_4$IrO$_4$; as for anisotropic exchange, we expect it to be a small relativistic correction to the isotropic exchange. On the other hand, we focus on the SIA, which is mainly determined by the metal center and its first coordination crystal field [43]. As a further confirmation of the magnitude of the SIA of a specified Ir ion, we replaced the three neighboring Ir$^{4+}$ ions with nonmagnetic Si$^{4+}$ ions in a supercell doubled along the crystallographic *c* axis. In this way, all other intersite NN or NNN magnetic exchange interactions vanish and the only contribution left is the SIA of the Ir$^{4+}$ ion. After checking that the crystal field splitting is unchanged with respect to the original configuration in Na$_4$IrO$_4$, our results (see Table III) show, as expected, a comparable magnitude of MCA and SIA energies.

Based on DFT electronic structure calculations for various spin-ordered magnetic insulating states, the spin exchange parameters can be obtained by mapping the relative energies of the magnetic ordered states onto Heisenberg or Ising Hamiltonian [44, 45, 46]. Generally, the spin Hamiltonian can be described by the classical Heisenberg model:

$$H = -\frac{1}{2}\sum_{i,j} J_{i,j} S_i \cdot S_j \qquad (7)$$

where $S_i$ represents a spin operator at site $i$ of the compound and the negative/positive value of $J$ denote AFM/FM interactions, respectively.

The spin exchange parameters $J_1$, $J_2$, and $J_3$ in Na$_4$IrO$_4$ are illustrated in Figure 1. Based on the optimized lattice parameters for the AFM unit cell as shown in Table I within GGA, we artificially construct five special magnetic ordering states (*i.e.*, FM, AFM1, AFM2, AFM3 and AFM4). The hypothetical FM state corresponds to a parallel alignment of all magnetic moments, whereas the other four AFM states are symmetry broken arrangements in the $\sqrt{2}\times\sqrt{2}\times 2$ supercell.

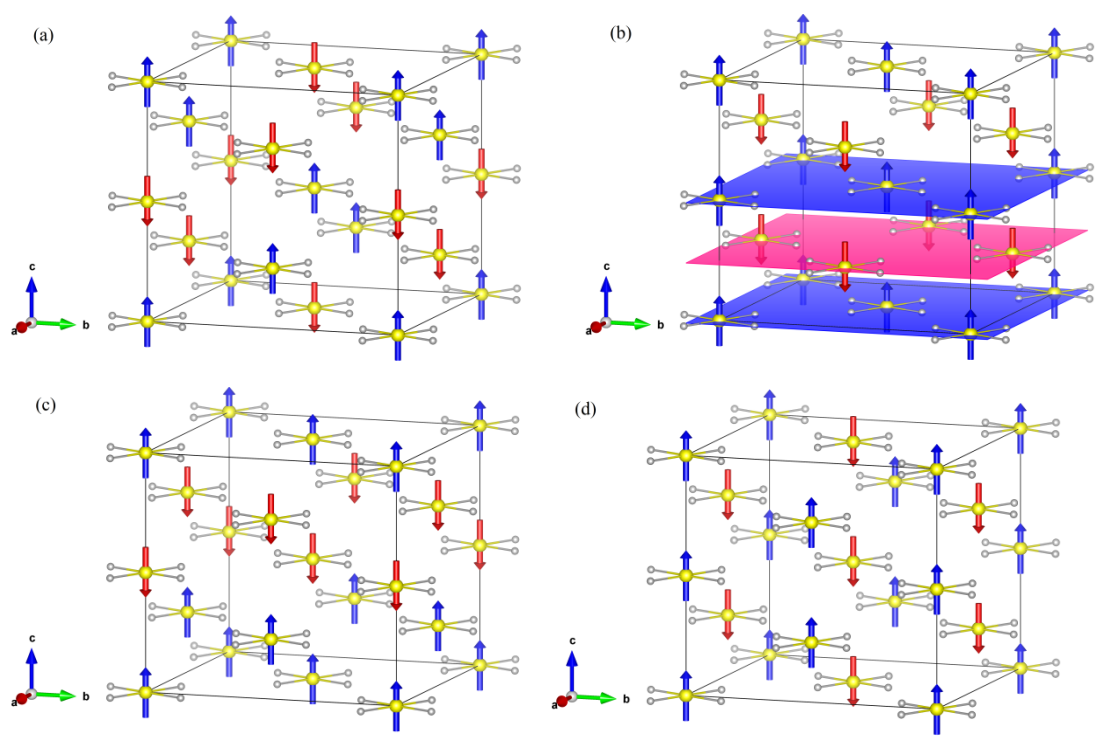

**Figure 6** Schematic representations of the four hypothetical AFM spin ordering arrangements in the $\sqrt{2}\times\sqrt{2}\times 2$ supercell (only IrO$_4$ square-plane are shown for clarity): (a) AFM1 (all magnetic moments are antiparallel to each other both in the *ab* plane and along *c* axis), (b) AFM2 (all magnetic moments are parallel to each other both in the *ab* plane and along *c* axis, but the NNN magnetic moments are antiparallel aligned), (c) AFM3 (all magnetic moments are antiparallel to each other along *c* axis, but parallel aligned in the *ab* plane ) and (d) AFM4 (all magnetic moments are parallel to each other along *c* axis, but antiparallel aligned in the *ab* plane). The big (yellow) and small (gray) spheres denote the Ir and O atoms; the up (down) arrows represent the magnetic moment orientations. The FM planes for the AFM2 magnetic ground state

are highlighted in (b), in order to better illustrate the stacking of FM planes antiferromagnetically coupled along the *c*-axis.

As shown by the DFT calculation results, the $Ir^{4+}$ ions are in intermediate-spin states with formal $S = 3/2$ in $Na_4IrO_4$ (see Figure 4(d)). In terms of exchange parameters, the spin exchange interaction energies (per f.u.) of the five magnetic ordering states are written as

$$E_{FM} = -\frac{9}{4}(J_1 + 4J_2 + 2J_3)$$

$$E_{AFM1} = -\frac{9}{4}(-J_1 - 2J_3)$$

$$E_{AFM2} = -\frac{9}{4}(J_1 - 4J_2 + 2J_3)$$

$$E_{AFM3} = -\frac{9}{4}(-J_1 + 2J_3)$$

$$E_{AFM4} = -\frac{9}{4}(J_1 - 2J_3)$$

(8)

Thus, by mapping the energy differences of these states in terms of the spin-exchange parameters with the corresponding energy differences from DFT calculations, we obtain

$$J_1 = -\frac{2}{9}(E_{AFM4} - E_{AFM1})$$

$$J_2 = \frac{1}{18}(E_{AFM2} - E_{FM})$$

$$J_3 = -\frac{1}{9}(E_{AFM3} - E_{AFM1})$$

(9)

**Table IV** Energy difference relative to the reference AFM2 state (meV/f. u.) and calculated spin exchange parameters (meV).

|  | AFM1 | AFM2 | AFM3 | AFM4 | FM | $J_1$ | $J_2$ | $J_3$ | $J_1/J_2$ |
| --- | --- | --- | --- | --- | --- | --- | --- | --- | --- |
| GGA | 9.21 | 0 | 9.05 | 11.33 | 23.74 | -0.47 | -1.32 | 0.02 | 0.36 |
| GGA + *U* | 4.21 | 0 | 4.09 | 6.16 | 12.61 | -0.43 | -0.70 | 0.01 | 0.62 |
| GGA + SOC | 8.48 | 0 | 8.27 | 10.50 | 21.93 | -0.45 | -1.22 | 0.02 | 0.37 |
| GGA + SOC + *U* | 3.93 | 0 | 3.79 | 5.89 | 11.98 | -0.43 | -0.67 | 0.02 | 0.65 |

Using the calculated energy values of the five magnetic ordering states, we obtain the spin-exchange parameters summarized in Table IV. SOC has a small impact on the spin exchange parameters, whereas the Coulomb interactions show remarkable influence, because the exchange coupling parameters $J$ are inversely proportional to the Hubbard $U$ [17]. The AFM2 state is the most stable, its total energy being lower than the other four magnetic states. The energies of the AFM1 state are comparable to the AFM3 state, reflecting very weak spin coupling interactions $J_3$ in the $ab$ plane. The negligible $J_3$ is consistent with the loosely connected structure and the large in-plane distances (about 7.2 Å) between the $Ir^{4+}$ ions along $a$ or $b$ axis. The other two spin exchange interactions $J_1$ and $J_2$ are AFM, and the NN interaction $J_1$ is smaller than the NNN interaction $J_2$, showing an inverse trend with respect to the distances for the NN (about 4.7 Å along $c$ axis) and NNN (about 5.6 Å along the diagonal of the unit cell) $Ir^{4+}$ ions. However, this is reasonable, when considering the unusual crystal structure of $Na_4IrO_4$, where a given Ir site has two NN and eight NNN coordination $Ir^{4+}$ ions. It should also be noted that both the NN and NNN interactions are AFM, so a geometrical magnetic frustration might arise in $Na_4IrO_4$, as the system cannot simultaneously satisfy all the NN and NNN AFM spin exchange interactions. However the large SIA favors the collinear alignment of the magnetic moments. Indeed, according to the experimental results, the frustration index $f = |\theta|/T_N$ is close to 3, and the calculated ratio of $J_1/J_2$ are far from 1 in all the cases (see Table IV), so a spin frustration does not occur, as confirmed by the AFM ordering obtained from magnetic susceptibility measurements [8].

Using the UppASD (Uppsala Atomistic Spin Dynamics) package [47], we perform MC simulations to capture the dynamical properties of the spin systems at finite temperatures for a 16 × 16 × 16 supercell based on the classical spin Hamiltonian [17]:

$$H = -\frac{1}{2}\sum_{i,j} J_{i,j} S_i \cdot S_j + \sum_i K S_{iz}^2 \qquad (10)$$

where the spin exchange parameters $J_{i,j}$ within GGA + SOC + $U$ are summarized in Table IV, while the SIA energy $KS_{iz}^2$ is given in Table III. To obtain the transition temperature $T_N$, we evaluate the order parameter (i.e. staggered magnetization related to the AFM2 magnetic configuration) and the specific heat at a given temperature $T$. As shown in Figure 7, without considering the SIA, the order parameter and the specific heat give similar results. The critical temperature $T_N$ is 28 K, evaluated from the peak position of the specific heat or from the values where the order parameter becomes negligible. The critical temperature as well the height of the specific heat peak increase upon including SIA in the MC simulations. The value of $T_N$ increases to 57 K, similar to the case in another infinite-layer oxide $SrFeO_2$ with square-planar coordination, where the critical temperature also increases when adding the magnetic anisotropy energy [17]. The magnetism is always collinear in $Na_4IrO_4$, considering the SIA, the spin moments being along the *c* axis perpendicular to the $IrO_4$ square-plane. In both cases, $Na_4IrO_4$ relaxes to the same AFM2 magnetic ground state with FM *ab* planes, antiferromagnetically coupled out-of-plane, corresponding to an antiparallel alignment of the spin magnetic moment of two Ir atoms in the crystallographic unit cell, revealed by first-principles calculations to be the lowest-energy magnetic state.

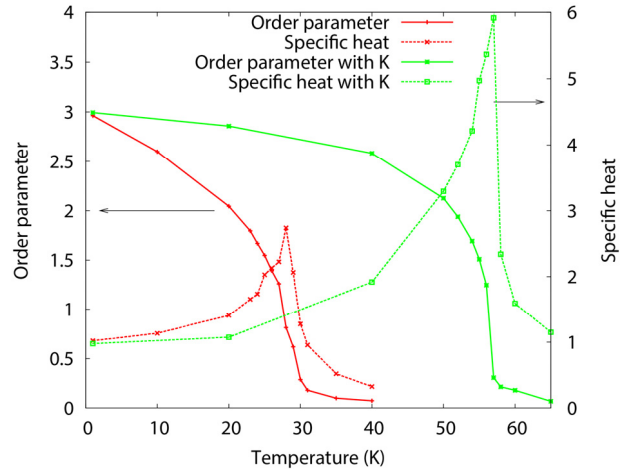

**Figure 7** Order parameter (solid line and left axis) and specific heat (dashed line and right axis) of $Na_4IrO_4$ calculated as a function of temperature on the basis of the classical spin Hamiltonian without SIA (red curves) and with SIA (green curves).

## IV. CONCLUSIONS

In summary, the novel square-planar coordination of $IrO_4$ plaquettes plays a crucial role in the electronic structure and magnetic properties of $Na_4IrO_4$, as shown by our comprehensive DFT calculations joint with MC simulations. The unusual square-planar crystal field and the strong hybridization effects give rise to an intermediate-spin state and to an insulating electronic structure, robust against different magnetic ordering, Coulomb parameters and even relativistic interactions. SOC produces a large MCA with an easy axis along the *c* axis perpendicular to the $IrO_4$ square-plane. When spin exchange interactions are evaluated by total energy calculations and mapping analysis, quite weak AFM interactions are obtained, consistent with the picture of rather isolated IrO$_4$ units. Moreover, MC simulations predict a quite low Néel temperature, consistent with experiments, and a collinear long-range AFM magnetic ground state. We hope our theoretical simulations will stimulate experimental works aimed at detailed magnetic properties measurements and characterizations, to further understand the magnetic ground state and exploit the large anisotropy of the uncommon square-planar coordinated Na$_4$IrO$_4$.

**ACKNOWLEDGMENTS**

We thank P. Barone, J. Hellsvik and K. Liu for useful discussions. This work was supported by the CARIPLO Foundation through the MAGISTER Project Rif. 2013-0726 and by IsC43 "C-MONAMI" Grant at Cineca Supercomputing Center. X. M. was sponsored by the China Scholarship Council (No. 201508420180), National Natural Science Foundation of China (No. 31600592), Scientific and Outstanding Young Science and Technology Innovation Team Program of Hubei Provincial Colleges and Universities (No. T201514).